\documentclass[12pt]{article}

\usepackage{hepth}
\usepackage{hyperref}
\usepackage{graphicx}

\begin{document}

\newcommand{\ie}{i.e.,\ }
\newcommand{\eg}{e.g.,\ }


\newcommand{\rmd}{\,\mathrm{d}}

\newcommand{\Tr}{\operatorname{tr}}

\newcommand{\idmat}{\mathbb{I}}

\newcommand{\bzero}{\mathbf{0}}

\newcommand{\re}{\operatorname{Re}}
\newcommand{\im}{\operatorname{Im}}

\newcommand{\e}[1]{\operatorname{e}^{#1}}

\newcommand{\bp}{\bar{\phi}}

\newcommand{\G}[2]{\mathcal{G}^{#1}_{\;\;#2}}
\newcommand{\R}{\mathcal{R}}

\newcommand{\mfa}{\mathfrak{a}}
\newcommand{\mfb}{\mathfrak{b}}
\newcommand{\mfc}{\mathfrak{c}}
\newcommand{\mfd}{\mathfrak{d}}
\newcommand{\mfe}{\mathfrak{e}}

\newcommand{\areg}{\mfa_{\mathrm{reg}}{}}
\newcommand{\asing}{\mfa_{\mathrm{sing}}{}}

\newcommand{\adom}{\hat{\mfa}}
\newcommand{\asub}{\check{\mfa}}

\newcommand{\sour}{\mathfrak{s}}
\newcommand{\resp}{\mathfrak{r}}

\newcommand{\vev}[1]{\left\langle{#1}\right\rangle}

\newcommand{\bra}[1]{\langle{#1}|}
\newcommand{\ket}[1]{|{#1}\rangle}

\providecommand{\op}{\mathcal{O}}

\newcommand{\Order}[1]{\mathcal{O}\left(#1\right)}

\newcommand{\rmI}{\operatorname{I}}
\newcommand{\rmJ}{\operatorname{J}}
\newcommand{\rmK}{\operatorname{K}}
\newcommand{\rmF}{\operatorname{F}}

\newcommand{\rmP}{\operatorname{P}}

\newcommand{\rmB}{\operatorname{B}}

\newcommand{\N}{\mathcal{N}}

\newcommand{\cL}{\mathcal{L}}

\newcommand{\matM}{\mathcal{M}}

\newcommand{\vecB}{\mathcal{B}}

\newcommand{\Son}{S_\text{on-sh}}

\newcommand{\VEV}{VEV}

\newcommand{\greens}{\operatorname{\mathbf{G}}}

\newcommand{\lad}{L}

\newcommand{\scale}{p}
\newcommand{\dscale}{\scale\frac{\rmd}{\rmd \scale}}

\newcommand{\UV}{\mathrm{UV}}
\newcommand{\IR}{\mathrm{IR}}

\preprint{NA-DSF-06/2010}
\title{Running Scaling Dimensions in Holographic Renormalization Group Flows}

\authors{Wolfgang M{\"u}ck\footnote{e-mail: \texttt{wolfgang.mueck@na.infn.it}}}

\institution{Naples}{Dipartimento di Scienze Fisiche, Universit\`a degli Studi di Napoli "Federico II" and \cr
INFN, Sezione di Napoli, Via Cintia, 80126 Napoli, Italy}

\abstract{Holographic renormalization group flows can be interpreted in terms of effective field theory. Based on such an interpretation, a formula for the running scaling dimensions of gauge-invariant operators along such flows is proposed. 
The formula is checked for some simple examples from the AdS/CFT correspondence, but can be applied also in non-AdS/non-CFT cases.
}

\date{June 2010}

\maketitle

%
%
%
%
\section{Introduction}
\label{HRflow}

In Quantum Field Theory (QFT), the term \emph{renormalization} is used in at least two different, although related contexts \cite{PS, Weinberg2}. In the first, it is part of a \emph{procedure} -- regularization and renormalization -- to extract meaningful, \ie finite physical quantities from formally infinite expressions, such as effective actions or correlation functions. In the second context, the \emph{renormalization group} (RG), it embodies the fact that the strengths of physical interactions depend on the (momentum) scale, at which they occur. Phenomenologically, this can be described by effective Lagrangians \cite{Wilson:1973jj, Polchinski:1983gv} with scale-dependent (running) coupling constants. 

Similarly, the term \emph{holographic renormalization}, which appears in the context of the gauge-gravity dualities, especially in the AdS/CFT correspondence \cite{Maldacena:1997re, Gubser:1998bc, Witten:1998qj}, is used either with an emphasis on the procedural or the physical aspects. On the one hand, the procedure of calculating correlation functions of gauge-invariant boundary operators from the bulk on-shell action, which is known as holographic renormalization, involves adding counter terms at a cut-off boundary and rescaling of the boundary values of the bulk fields, in analogy with the QFT procedure \cite{deHaro:2000xn, Bianchi:2001kw, Skenderis:2002wp, Martelli:2002sp, Papadimitriou:2004ap, Papadimitriou:2004rz}. On the other hand, the bulk scalar fields can be interpreted as scale-dependent coupling constants. This allows to write down, in analogy to QFT, holographic beta functions and holographic versions of the Callan-Symanzik equation \cite{Akhmedov:1998vf, Alvarez:1998wr, Kehagias:1999tr, deBoer:1999xf, Li:2000ec, Erdmenger:2001ja}. This interpretation is particularly evident in the context of holographic RG flows -- bulk domain wall configurations, which are the gravity duals of RG flows in QFT and imply a $c$-theorem \cite{Girardello:1998pd, Porrati:1999ew, Freedman:1999gp, Anselmi:2000fu, Myers:2010xs} similar to Zamolodchikov's.

In this note, we shall consider holographic renormalization from the viewpoint of effective Lagrangians, identifying literally the bulk scalar fields with the running coupling constants of an effective field theory. More specifically, we consider holographic RG flows, which are identified as the bulk duals of particular solutions of the QFT RG equations. Identifying the bulk fluctuations around these backgrounds with (small) perturbations of the RG flow, we are able to derive a formula that captures the running of the dimension matrix of the QFT operators along the flow. The dimension formula \eqref{HRflow:Delta.holo} will be our main result. The new proposal we make to obtain it is to relate both, the radial variable \emph{and} the boundary momentum, to the QFT momentum scale. As a check, we present some examples from AdS/CFT. 

The dimension formula \eqref{HRflow:Delta.holo} solves a small puzzle concerning the running of operator dimensions along holographic RG flows. Given, for example, an RG flow between two distinct fixed points, the dimensions of the operators flow from their (bare) UV values to some different values in the IR. The intermediate flow is, of course, scheme dependent, but there should exist a smooth function, which interpolates between the UV and the IR dimensions. In AdS/CFT, no general function is known that would give such an  interpolation in all holographic RG flows. The holographic (anomalous) scaling dimension introduced in \cite{deBoer:1999xf} would be an interesting candidate, but there are several reasons why it fails in many cases. We shall give more details in Sec.~\ref{bulk}.

Although, in the present paper, the focus is primarily on holographic RG flows in AdS/CFT, the dimension formula readily generalizes to non-AdS/non-CFT cases, because the formalism that we use to treat the bulk fluctuations is independent of the existence of a conformal fixed point. Hence, just as the procedure of holographic renormalization can be applied in non-AdS/non-CFT settings \cite{Aharony:2005zr, Borodatchenkova:2008fw}, there is an interpretation of holographic RG flows without a UV fixed point in terms of effective Lagrangians. We leave the application to specific examples of non-AdS/non-CFT for future work. Similarly, we believe that the results are relevant for applications of AdS/CFT to condensed matter physics. 

Let us outline the rest of the paper. In Sec.~\ref{running}, we consider an RG flow in QFT and derive a formula for the matrix of running scaling dimensions along the flow. The bulk description of the RG flow and the holographic version of the dimension formula are given in Sec.~\ref{bulk}, where we also review and discuss the dimension formula of \cite{deBoer:1999xf}. In Sec.~\ref{examples}, the new dimension formula is applied to several examples from AdS/CFT. In particular, the flows interpolating between UV and IR fixed points are non-trivial checks of the formula.

%
%
\section{Running Scaling Dimensions in QFT}
\label{running}

To start the discussion, let us consider an effective Lagrangian, which contains some operators $\op_a$, with couplings $\phi^a$\footnote{Our notation, which is somewhat unusual for QFT, anticipates the bulk description.}
\begin{equation}
\label{HRflow:op.insertion}
	\int \rmd^d x\, \phi^a\, \op_a~.
\end{equation}
The couplings $\phi^a$ are running coupling constants depending on the momentum scale, $\scale$, by the renormalization group equation\footnote{For a textbook presentation, see Sec.~12.3 of \cite{PS}.}
\begin{equation}
\label{HRflow:coupling.flow}
	\dscale \phi^a = \beta^a(\phi)~,
\end{equation}
where the beta-functions $\beta^a$ form a vector in the space of couplings. Generally in perturbative QFT, $\beta^a$ is not known throughout the space of couplings, but only as a power series in the vicinity of fixed points. The same is true, in general, in AdS/CFT, since for example, the holographic beta function introduced in \cite{deBoer:1999xf} captures just the flow of the couplings due to the divergent terms of the bulk on-shell action. 

Let us consider, however, the case in which a particular, \emph{exact} solution of \eqref{HRflow:coupling.flow} is known, which we denote by $\bp(\scale)$. Then, we can study small deviations from this exact solution by linearizing \eqref{HRflow:coupling.flow} around $\bp(\scale)$. This yields
\begin{equation}
\label{HRflow:coupling.flow.lin}
	\dscale \mfa^a = \left[D_b \beta^a(\bp)\right] \mfa^b~,
\end{equation}
where $D_b$ denotes a covariant derivative in the space of couplings (provided there is a metric), and the $\mfa^a$ parameterize the \emph{linear} deviations of the couplings from $\bp(\scale)$. The derivative $D_b \beta^a(\bp)$ is related to the (scale-dependent) dimension matrix by
\begin{equation}
\label{HRflow:delta.def}
	D_b \beta^a(\bp) = [\Delta(\bp) -d]^a{}_b~.
\end{equation}
Hence, \eqref{HRflow:coupling.flow.lin} gives the well known relation \cite{PS} 
\begin{equation}
\label{HRflow:running2}
	\dscale \mfa^a + (d-\Delta)^a{}_b\, \mfa^b = 0~.
\end{equation}

The solution of the linearized flow equation \eqref{HRflow:running2} can be written in the form 
\begin{equation}
\label{HRflow:run.sol}
	\mfa^a(\scale) = \mfa^a{}_b(\scale)\, \mfa_{\text{ren}}^b~, \qquad \mfa^a{}_b(M)=\delta^a_b~,
\end{equation}
where the initial values $\mfa_{\text{ren}}^a$ are the \emph{renormalized couplings} at a renormalization scale $M$. Since the $\mfa_{\text{ren}}^a$ can be chosen arbitrarily, \eqref{HRflow:running2} gives rise to the matrix equation
\begin{equation}
\label{HRflow:running3}
	\dscale \mfa^a{}_c(\scale) + (d-\Delta)^a{}_b\, \mfa^b{}_c(\scale) = 0~.
\end{equation}
Hence, assuming that the flow is invertible, \ie that the inverse of the matrix $\mfa^a{}_b$ exists, one immediately obtains the dimension matrix along the exact flow $\bp(\scale)$,
\begin{equation}
\label{HRflow:Delta}
	\Delta^a{}_b(\scale) = d \delta^a_b + \left[\dscale \mfa^a{}_c(\scale)\right] \left[\mfa^{-1}(\scale)\right]^c{}_b~.
\end{equation}
%

%
%
%
\section{Holographic Dimension Formula}
\label{bulk}

Formula \eqref{HRflow:Delta} has a very nice holographic analogue, as our notation already suggests. To derive it, we consider a truncated (fake) SUGRA system with an action of the form
\begin{equation}
 \label{bdyn:action}
  S= \int \rmd^{d+1}x \sqrt{g} \left[ -\frac14 R +\frac12 G_{ab}\,g^{MN}\, \partial_M \phi^a \partial_N \phi^b +V(\phi)\right]~,
\end{equation}
where $M,N=0,\ldots d$, and $G_{ab}$ is a sigma-model metric on the space of scalar fields. The potential $V(\phi)$ is given in terms of a superpotential $W(\phi)$, 
\begin{equation}
 \label{bdyn:potential}
  V(\phi) = \frac12 G^{ab} W_a W_b -\frac{d}{d-1} W^2~,
\end{equation}
with $W_a=\partial W/\partial \phi^a$. This ensures the existence of holographic RG flows --- BPS Poincar\'e-sliced domain wall solutions --- which may be written in the form 
\begin{equation}
\label{HRflow:bg}
  \rmd s^2 = \frac{(d-1)^2}{4[W(\bp)]^2} \rmd \sigma^2 + \e{2\sigma} \eta_{\mu\nu}\rmd x^\mu \rmd x^\nu~, \qquad
  \partial_\sigma \bp^a = -\frac{d-1}2 \frac{W^a(\bp)}{W(\bp)}~,
\end{equation}
where $\mu,\nu=1,\ldots d$. The radial variable $\sigma$, which we use throughout this paper, is just the warp function, $\sigma=A(r)$, of the more conventional form \cite{Freedman:1999gp}.

Fluctuations around the background \eqref{HRflow:bg} are described in a gauge-invariant fashion \cite{Bianchi:2003ug, Berg:2005pd} by a scalar field vector $\mfa^a$, which satisfies the linearized equation of motion
\begin{equation}
\label{HRflow:eom}
	\left[ \left( D_\sigma +\matM +d -\frac2{d-1} \vecB_a \vecB^a \right) \left( D_\sigma -\matM \right) -
	\frac{(d-1)^2}{4W^2} \e{-2\sigma} k^2 \right] \mfa(\sigma,k) = 0~.
\end{equation}
For brevity, we have omitted the vector indices, and $k^\mu$ ($k=\sqrt{k_\mu k^\mu}$) denotes the boundary momentum, \ie the conjugate of the coordinates $x^\mu$ in \eqref{HRflow:bg}. The vector $\vecB$ and the matrix $\matM$ are given by 
\begin{equation}
\label{HRflow:matM}
	\vecB_a = -\frac{d-1}2 \frac{W_a}W~,\qquad \matM^a{}_b = D_b \vecB^a~.
\end{equation}
In \eqref{HRflow:eom}, $W$, $\vecB$ and $\matM$ are evaluated on the background \eqref{HRflow:bg} and are, therefore, functions of $\sigma$. $D_a$ is the covariant derivative in field space, defined from the metric $G_{ab}$ as usual, $D_a W^b = \partial_a W^b +\G{b}{ac}W^c$, with Christoffel symbols $\G{b}{ac}$. Similarly, $D_\sigma$ is the background covariant derivative 
\begin{equation}
\label{HRflow:D.def}
	D_\sigma \mfa^a = \partial_\sigma \mfa^a + \G{a}{bc} (\partial_\sigma \bp^b) \mfa^c~.
\end{equation}

Typically, a regularity condition in the bulk interior allows for $n$ regular solutions of \eqref{HRflow:eom}, if one has $n$ scalar fields, \ie if $\mfa$ has $n$ components. In addition to $\mfa^a$, the linearized bulk fields comprise the traceless transverse components of the metric, which satisfy a similar equation of motion, but we shall not need them.

Let us now derive the holographic analogue of \eqref{HRflow:Delta}. To do this, we consider the bulk background configuration \eqref{HRflow:bg} as the holographic dual of an exact solution of the RG equations \eqref{HRflow:coupling.flow}, with the scalars $\bp(\sigma)$ representing the running couplings, $\bp(\scale)$. The QFT, which implements this flow, is perturbed by the insertion of operators $\op_a$ with small couplings $\mfa^a$, which exhibit some running on their own. As our notation already suggests, we are identifying the \emph{linearized} bulk fluctuations $\mfa^a$ with the \emph{linearized} QFT couplings of the previous section. To do this, we also need to specify how the QFT momentum scale $\scale$ is related to the bulk coordinates, on which the bulk field $\mfa$ depens. Usually, one relates the radial variable to the QFT scale, which allows to give a local interpretation to the RG flow \cite{Erdmenger:2001ja}, but we propose that also the momentum $k$ should be fixed. Indeed, $k$ is a Euclidean momentum scale, just as $\scale$ is in QFT. Hence, let us first identify the (Euclidean) momentum $k$ of the bulk description with the QFT scale, $k=\scale$.  
Then, as the QFT coupling depends only on $\scale$, we also need to fix the radial variable $\sigma$ in terms of $\scale$. The most natural relation follows from considering the variable transformation
\begin{equation}
\label{HRflow:shift}
	\sigma \to \sigma'=\sigma-\delta\sigma~,\qquad x^\mu \to x'{}^\mu=\e{\delta\sigma} x^\mu~,
\end{equation}
which leaves the $d$-dimensional part of the background metric \eqref{HRflow:bg} invariant and acts as a scale transformation on the boundary. Therefore, we set  
\begin{equation}
\label{HRflow:scale}
	\scale = k \qquad \text{and}\qquad \scale = M\e{\sigma}~,
\end{equation}
where $M$ is a renormalization scale, which introduces scheme dependence into the game. 

As a last detail, we should implement the initial conditions of \eqref{HRflow:run.sol} at the renormalization scale $M$. If we denote by $\mfa_i^a$ a set of independent \emph{regular} bulk fields (specified by the index $i$), normalized in such a way that the asymptotically dominant term is always independent of $k$, just as in AdS/CFT, then we can identify the QFT and the bulk descriptions of the running couplings by
\begin{equation}
\label{HRflow:lambda.ident}
	\mfa^a{}_b(\scale) \equiv \mfa^a_i(\sigma, k)\, [\mfa^{-1}(0,M)]^i_b~. 
\end{equation}
$(\mfa^{-1})^i_a$ is the matrix inverse of $\mfa^a_i$, but its presence is irrelevant for what follows. As a bonus, \eqref{HRflow:lambda.ident} determines also how, in the procedure of holographic renormalization, the boundary values of the bulk fluctuations at the cut-off boundary should be related to the renormalized couplings. 

Finally, let us consider the evolution of the couplings with the scale $\scale$. Because both, $k$ and $\sigma$, are related to the scale $\scale$, and because $\mfa^a$ is a field vector, the we should write
\begin{equation}
\label{HRflow:flow.def}
	\dscale \mfa^a{}_b \equiv \left[\left( D_\sigma +k\partial_k\right) \mfa^a_i(\sigma,k)\right] 
	\left[\mfa^{-1}(0,M)\right]^i_b~.
\end{equation}
Putting everything together, we obtain the holographic analogue of \eqref{HRflow:Delta} as
\begin{equation}
\label{HRflow:Delta.holo}
	\Delta^a{}_b = d \delta^a_b + \left[(D_\sigma +k\partial_k)\mfa^a_i(\sigma,k)\right] [\mfa^{-1}(\sigma,k)]^i_b~.
\end{equation}
This is our main result. Let us note that \eqref{HRflow:Delta.holo} is independent of the choice of regular basis solutions, because the index $i$ is traced over. Morever, the matrix $[\mfa^{-1}(0,M)]^i_b$, which we introduced in \eqref{HRflow:lambda.ident} to satisfy the renormalization conditions, has dropped out. Equivalently, we may consider the scale dependence of $\mfa$ after substituting $k=M\e{\sigma}$ into the bulk fields. Then, \eqref{HRflow:Delta.holo} transforms into a similar formula without the derivative with respect to $k$.

In \cite{deBoer:1999xf}, formulae for the holographic beta function and holographic scaling dimensions of operators were given.\footnote{Often, one distinguishes between the bare and the anomalous dimensions, $\Delta_0$ and $\gamma$, respectively. In \cite{deBoer:1999xf}, $\Delta_0=d$ is assumed, and $\gamma$ is defined holographically. Here, we shall just talk about the full dimension $\Delta=\Delta_0+\gamma$, which is of course scale-dependent.} 
Let us compare them with our results. With our slightly different conventions, the formulae of \cite{deBoer:1999xf} read
\begin{equation}
\label{HRflow:hol.beta}
	\beta^a(\phi) = -\frac{d-1}2 \frac{U^a(\phi)}{U(\phi)}~,
\end{equation}
and
\begin{equation}
\label{HRflow:hol.Delta}
	\Delta^a{}_b(\phi) = d\delta^a_b + D_b \beta^a(\phi)~.
\end{equation}
They are defined in terms of a potential $U(\phi)$ that satisfies the same equation \eqref{bdyn:potential} as the superpotential $W(\phi)$ does. So, if one could identify $U$ with $W$, one would obtain $\beta^a=\partial_\sigma \bp^a$ and $\Delta = d+\matM$ along the holographic RG flow. 
However, in the general case, there are important differences between the two potentials, which stem from the nonlinearity of \eqref{bdyn:potential}. Let us consider AdS/CFT for simplicity. Whereas $W$ is given from the start, $U$ is practically defined as a power series in the fields around the AdS fixed point. The coefficients of the quadratic terms satisfy a quadratic equation, which has obviously two solutions. As for the counter terms in holographic renormalization \cite{Martelli:2002sp, Papadimitriou:2004ap}, only one of the two solutions is adequate to capture the asymptotic behaviour of the bulk fields \cite{deBoer:1999xf}. Hence, it may happen that the power series expansion of $W$ differs from the expansion of $U$ in the quadratic terms, which determine the value of the dimension at the fixed point. If this is the case for certain fields, then it is easy to show that, for the operators dual to these fields, $\Delta=-\matM$ at the fixed point, where by $\matM$ we intend the appropriate eigenvalue of the matrix. Therefore, one may be tempted to say that either $d+\matM$ or $-\matM$ provides the operator dimension along the flow, but this is not correct, either. In fact, for a holographic RG flow between two fixed points, one has to make the appropriate choice (for each eigenvalue of the matrix) at each fixed point, and it is not guaranteed that the two choices agree. One of the examples we provide later represents just such a case. 

In addition, the order-by-order relations for the coefficients of $U(\phi)$ break down for the \emph{finite} terms (of total dimension $d$). If there are fields with different quadratic coefficients in $U$ and $W$, the breakdown typically implies that \eqref{bdyn:potential} cannot be solved as a power series expansion, and one obtains contributions to the holographic Weyl anomaly \cite{Henningson:1998gx, Martelli:2002sp}. But even if all quadratic terms are equal, the finite terms of $U$ are ambiguous. Hence, in any case, the breakdown of \eqref{bdyn:potential} makes it impossible to give significance to $U$ beyond the divergent terms, so that \eqref{HRflow:hol.beta} and \eqref{HRflow:hol.Delta} should be considered as valid only in the UV, just as 1-loop corrections.

%
\section{Examples}
\label{examples}

%
%

\paragraph{AdS -- QFT at a Fixed Point}
As a first check of the dimension formula \eqref{HRflow:Delta.holo}, let us consider a pure AdS bulk background, which is dual to some conformal field theory (CFT). Hence, there is no scale dependence, and \eqref{HRflow:Delta.holo} should reflect this.
Massive scalar fields in the bulk are dual to gauge-invariant operators of certain scaling dimensions, $\Delta_i= d/2 +\alpha_i$, were the $\alpha_i$ are related to the bulk masses. The regular bulk solutions of \eqref{HRflow:eom}, normalized such that the asymptotically leading term is independent of $k$, are given by 
\begin{equation}
\label{ex:AdS.reg}
  \mfa_i^a(z,k) = \delta^a_i \frac{2}{\Gamma(\alpha_a)} \left( \frac{k}{2} \right)^{\alpha_a} 
     z^{d/2} \rmK_{\alpha_a} (k z)~,
\end{equation}
where $z=\e{-\sigma}$, and the $\rmK_\alpha$ are modified Bessel functions. 

To describe the scale dependence, it is easiest to eliminate $k$ by the relation $kz=M$, which gives rise to (the $k$-independent coefficients are irrelevant here)
\begin{equation}
\label{ex:AdS.reg.M}
	\mfa^a_i (z) \sim \delta^a_i\, z^{-(d/2-\alpha_i)} M^{\alpha_i} \rmK_{\alpha_i} (M)~.
\end{equation}
As mentioned in Sec.~\ref{bulk}, from this expression one may read off the renormalization of the boundary values at the cut-off boundary (at $z=z_c$) in terms of the renormalized couplings. As expected, one just gets a factor $z_c^{-(d/2-\alpha_i)}$, so that the renormalized couplings coincide with the regular AdS boundary conditions of Breitenlohner and Freedman \cite{Breitenlohner:1982jf}.\footnote{It  would be interesting to consider also the irregular boundary conditions \cite{Klebanov:1999tb, Mueck:1999kk}.}

Substituting \eqref{ex:AdS.reg.M} into the dimension formula \eqref{HRflow:Delta.holo} immediately leads to the expected result
\begin{equation}
\label{ex:AdS.dim}
	\Delta^a{}_b = \left(\frac{d}2 +\alpha_a\right) \delta^a_b~.
\end{equation}

%
%

\paragraph{Holographic RG Flows Between Two Fixed Points}
To have a non-trivial check of the dimension formula, we consider two examples of RG flows that end at an IR fixed point.
Specifically, we consider the $SU(2)\times U(1)$ flow in $d=4$ \cite{Freedman:1999gp,Pilch:2000fu}, which is the gravity dual of the Leigh-Strassler flow, and its $d=3$ cousin, the $SU(3)\times U(1)$ flow \cite{Ahn:2000aq,Ahn:2000mf,Corrado:2001nv}. In both cases, the holographic RG flow involves two (active) scalar fields and is known only numerically. The operator dimensions at the fixed points, though, are analytically known, as they can be calculated from the holographic formula \eqref{HRflow:hol.Delta}. 
For the $SU(2)\times U(1)$ flow, one has
\begin{equation}
\label{ex:su2.dims}
	\Delta_\UV = (3,2)~,\qquad \Delta_\IR = (1+ \sqrt{7}, 3+\sqrt{7})~.
\end{equation}
For the $SU(3)\times U(1)$ flow, the dimensions at the fixed points are
\begin{equation}
\label{ex:su3.dims}
	\Delta_\UV = (2,2)~,\qquad \Delta_\IR = \left(\frac12 + \frac12\sqrt{17}, \frac52 +\frac12\sqrt{17}\right)~.
\end{equation}
For more information on the two flows, we refer to \cite{Freedman:1999gp,Pilch:2000fu,Ahn:2000aq,Ahn:2000mf,Corrado:2001nv}. 
It has been confirmed numerically, by a calculation of the spectral functions, that the operator dimensions interpolate smoothly between the respective $\UV$ and $\IR$ values of \eqref{ex:su2.dims} and \eqref{ex:su3.dims} \cite{Mueck:2008gv}. 

\begin{figure}[th]
\includegraphics[width=0.5\textwidth]{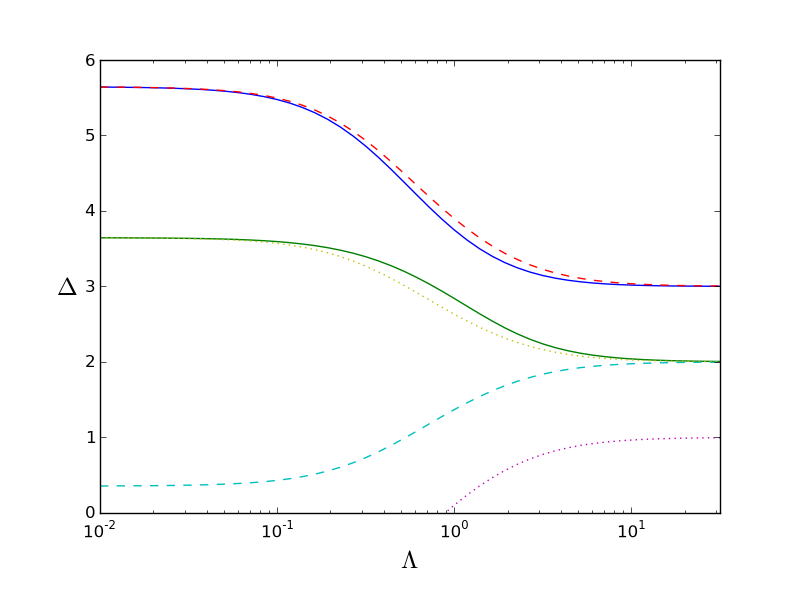}%
\hfill%
\includegraphics[width=0.5\textwidth]{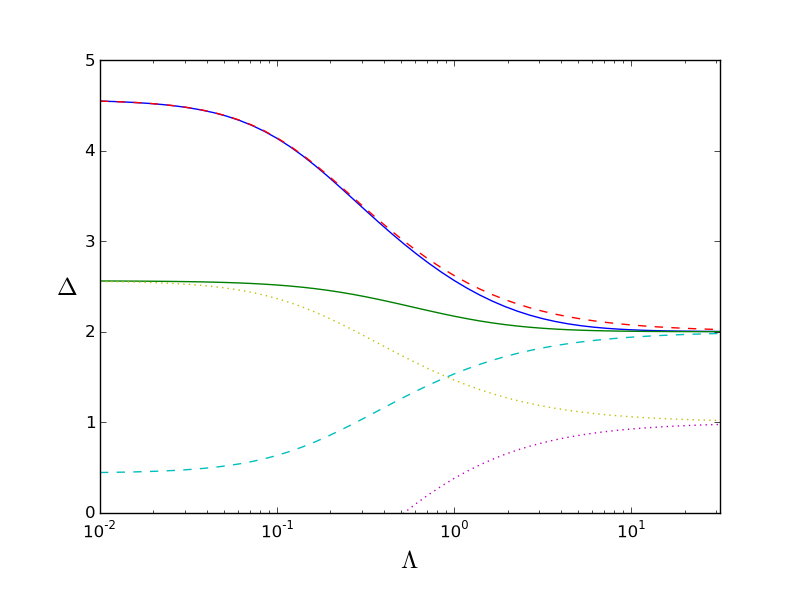}%
\caption{Scaling dimensions of the operators in the $SU(2)\times U(1)$ (left) and the $SU(3)\times U(1)$ flows (right). Shown are the eigenvalues of the dimension matrix \eqref{HRflow:Delta.holo} (solid lines), the eigenvalues of $d+\matM$ (dashed lines) and of $-\matM$ (dotted lines). The renormalization scale $M$ has been fixed to some convenient value. \label{ex:SUdims.fig}}
\end{figure}

We have evaluated numerically, for the $SU(2)\times U(1)$ and the $SU(3)\times U(1)$ flows, the dimension formula \eqref{HRflow:Delta.holo}.\footnote{The \texttt{Sagemath} worksheets \cite{sage} are available from the author.} The results are shown in Fig.~\ref{ex:SUdims.fig} and show very nicely that the eigenvalues of \eqref{HRflow:Delta.holo} indeed interpolate between the correct fixed point values. The eigenvalues of the matrices $d+\matM$ and $-\matM$ are also shown, which are the naive candidates for the operator dimensions, as mentioned in Sec.~\ref{bulk}. The fixed point values must agree with one of the naive choices, which is clearly the case. Moreover, in the $SU(2)\times U(1)$ case, one is lucky that $\Delta=4-\Delta$ for $\Delta=2$, so that switching between the two naive functions does not change that particular UV value, and one may choose the one that gives the correct IR value. However, this is clearly not possible in the $SU(3)\times U(1)$ case, where one can only obtain either the correct UV or the correct IR value.

%
%

\paragraph{GPPZ flow}
In order to demonstrate that the dimension formula \eqref{HRflow:Delta.holo} can also be applied in the absence of an IR fixed point, let us consider the dimensions of the active and the inert scalars in the GPPZ flow \cite{Girardello:1998pd}. We refer the reader to \cite{Muck:2004qg} for the details on the background and fluctuation dynamics. 

Both operators have bare dimensions $\Delta_\UV=3$. Of course, as there is no fixed point in the IR, we do not have an independent confirmation of their IR dimensions, and we cannot obtain them from the spectrum either, because it is discrete \cite{Muck:2001cy, Bianchi:2001de}. Hence, our formula should give a prediction for the IR dimensions, as we cannot trust the naive matrices $d+\matM$ and $-\matM$ away from fixed points.

\begin{figure}[th]
\includegraphics[width=0.5\textwidth]{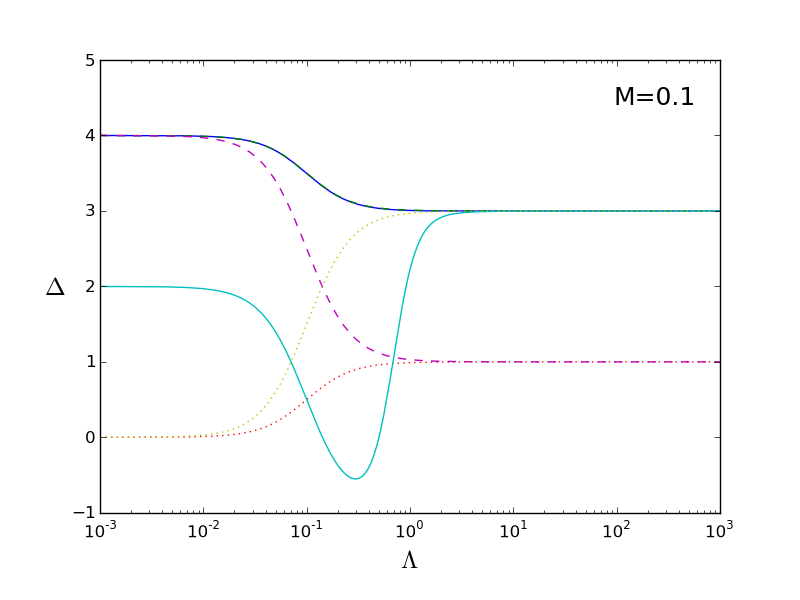}%
\includegraphics[width=0.5\textwidth]{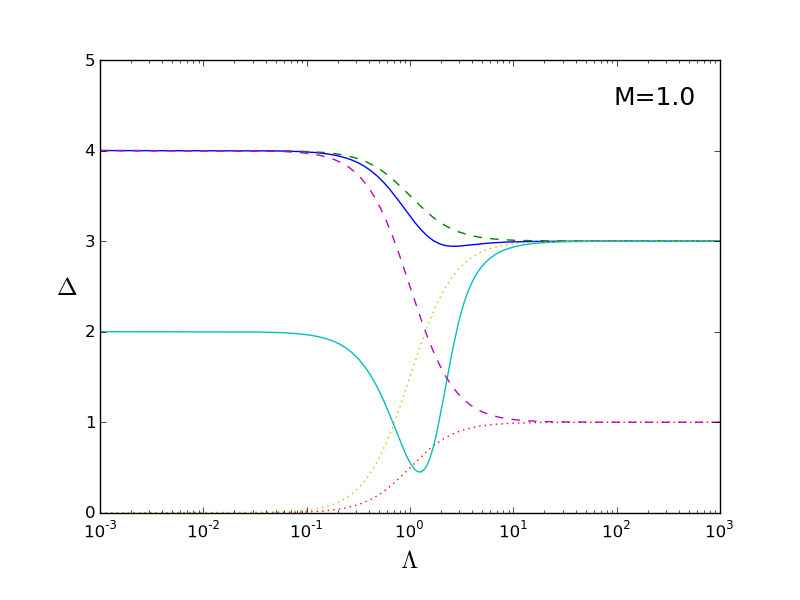}\\%
\includegraphics[width=0.5\textwidth]{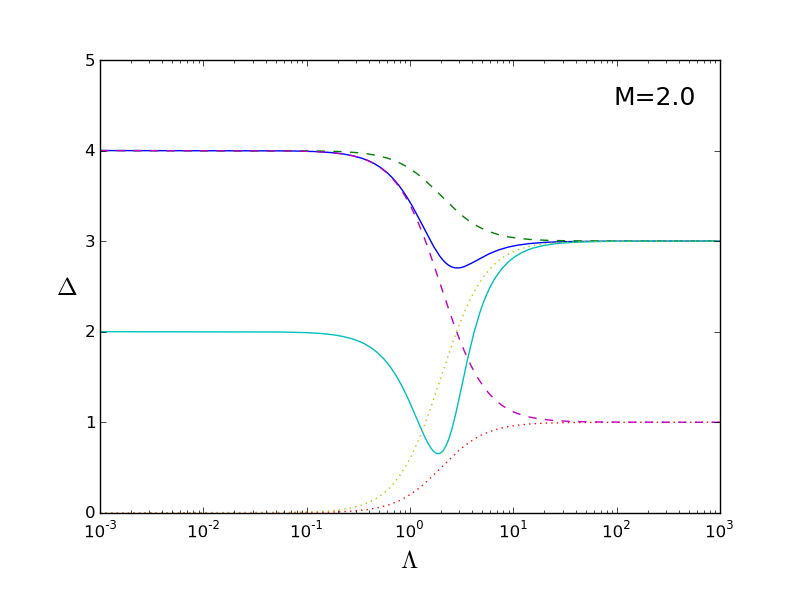}%
\includegraphics[width=0.5\textwidth]{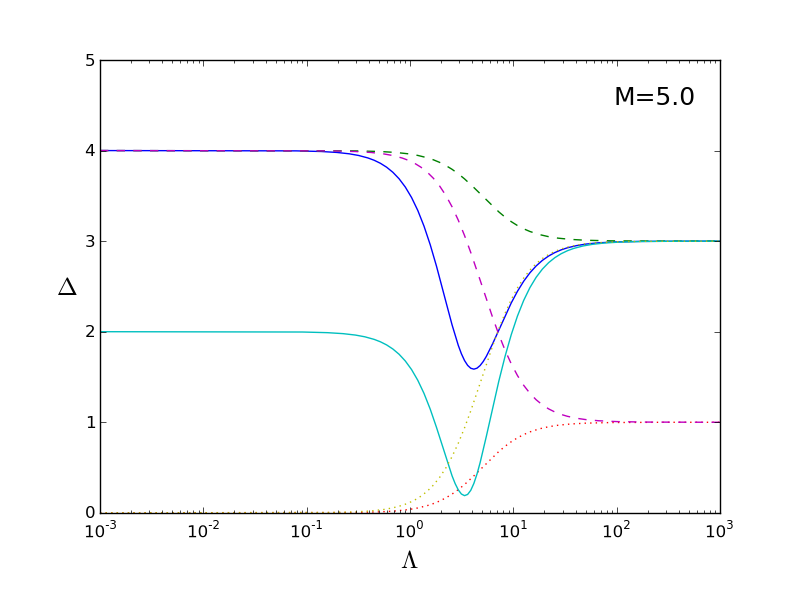}%
\caption{Running dimensions of the operators in the GPPZ flow. The solid, dashed and dotted lines represent the eigenvalues of the dimension matrix \eqref{HRflow:Delta.holo}, of $d+\matM$ and of $-\matM$, respectively. For $M=0.1$, the dimension of $\op_\phi$ virtually agrees with the larger eigenvalue of $d+\matM$. \label{ex:GPPZ.dims}}
\end{figure}

Let us denote the active and inert scalar fields by $\phi$ and $\chi$, respectively.\footnote{We choose $\chi$ instead of $\sigma$, which is used in the literature, in order to avoid confusion with the radial coordinate.} The solutions of the gauge-invariant equations of motion for their fluctuations are \cite{Muck:2004qg}\footnote{The two components of the fluctuation vector are not coupled.}
\begin{equation}
\label{ex:GPPZ.sol}
	\mfa^\phi = \Gamma\left(\frac{3+\alpha}2\right)\Gamma\left(\frac{3-\alpha}2\right) 
	\sqrt{1-u} \rmF\left(\frac{1+\alpha}2,\frac{1-\alpha}2;2;u\right)
\end{equation}
and
\begin{equation}
\label{ex:GPPZ.sol2}
	\mfa^\chi = \Gamma\left(\frac{3+\beta}2\right)\Gamma\left(\frac{3-\beta}2\right) 
	\sqrt{1-u} \rmF\left(\frac{1+\beta}2,\frac{1-\beta}2;2;u\right)~, 
\end{equation}
respectively. Here, $\rmF$ denotes a Gauss' hypergeometric function, and the coefficients are $\alpha=\sqrt{1-k^2}$ and $\beta=\sqrt{9-k^2}$. The radial variable $u$ is related to $\sigma$ by $\e{2\sigma}=u/(1-u)$. 

Applying \eqref{HRflow:Delta.holo} will lead to derivatives of the hypergeometric functions not just with respect to $u$, but also with respect to their parameters. Although it is possible to express those derivatives in terms of some generalized hypergeometric series \cite{Ancarani:2009zz}, the result would not be very enlightening. It is much easier to set $u=\scale^2/(M^2+\scale^2)$ and $k^2 =\scale^2$ and do the derivative with respect to $\scale$ numerically. We have performed this computation using the \texttt{mpmath} package that is included in \texttt{Sagemath} \cite{mpmath,sage}. The results for both scalars are shown in Fig.~\ref{ex:GPPZ.dims}. In order to illustrate the scheme dependence, several values of $M$ have been considered. It appears that, for certain values of $M$, an eigenvalue of $d+\matM$ or $-\matM$ can provide, either in the IR or the UV regime, a good approximation for the dimension of $\op_\phi$, but not of $\op_\chi$. In addition, the IR dimension of $\op_\chi$, according to our dimension formula, is 2, which is not reproduced by the naive functions.

%


\section*{Acknowledgments}
I would like to thank M.~Haack for stimulating discussions and his collaboration on a related project, as well as J.~Erdmenger for critical comments.
This work was partially supported by the INFN research initiative TV12, as well as by the Italian Ministry of Education and Research (MIUR), PRIN project 2005-023102. 


\begin{singlespace}
\bibliographystyle{JHEP}
\bibliography{holo_dims}
\end{singlespace}
\end{document}